\documentclass[3p,12pt,times,twocolumn,number,sort&compress]{elsarticle}
\usepackage[utf8]{inputenc} \usepackage{amsmath,amssymb} \usepackage{tabu}
\usepackage[outdir=./]{epstopdf}

\usepackage{color}
\usepackage[modulo]{lineno}
\usepackage{graphicx}

\usepackage[superscript,biblabel]{cite}

\graphicspath{{../Figures-1-2-3-etc/}}
             
\usepackage{amssymb} \usepackage[modulo]{lineno}
\usepackage[figuresright]{rotating}

\usepackage{amssymb} \usepackage[modulo]{lineno}
\usepackage[figuresright]{rotating}
\makeatletter
\def\ps@pprintTitle{%
	\let\@oddhead\@empty
	\let\@evenhead\@empty
	\def\@oddfoot{}%
	\let\@evenfoot\@oddfoot}
\makeatother
\begin{document}

\begin{frontmatter}


\title{Cross-location wind speed forecasting for wind energy applications using machine learning based models}

\author[ku,AN]{Valsaraj Perumpalot} \ead{valsaraj@anert.in} 
\author[ku]{G. V. Drisya} \ead{drisyavictoria@gmail.com} 
\author[ku]{K. Satheesh Kumar\corref{cor1}}
\ead{kskumar@keralauniversity.ac.in} \cortext[cor1]{Corresponding author}
\address[ku]{Department of Futures Studies, University of Kerala,
	Kariavattom, Kerala, India - 695~581} \address[AN]{Agency for Non-conventional Energy and Rural Technology (ANERT), Thiruvananthapuram, Kerala, India - 695~033}

\begin{abstract} The widespread utilisation of grid-integrated wind electricity necessitates accurate and reliable wind speed forecasting to ensure stable grid and quality power. Machine learning algorithm based wind speed forecasting models are getting increased attention in the literature owing to its superior ability to learn by effectively capturing the changing patterns from the data. Most of the reported wind forecasting models built on machine learning algorithms are location specific and tested against data adjacent to the training data.  In this work, we develop the machine learning based wind speed forecasting models and analyse their performance when applied to data from different cross- locations up to a year ahead. Two distinct machine learning models based on Support Vector Machine (SVM) and Random Forest (RF) algorithms have been developed and tested separately for a relatively large geographical area. The results of analysis of 1-hour forecasts obtained at various cross-locations and points of time up to a year ahead show 80\% of predictions within a Root Mean Square Error (RMSE) of 1.5 m/s, 95\% within 2.5 m/s and 98\% within an RMSE of 3.5 m/s. The 75\% of 2-hour predictions are within RMSE of 1.5 m/s, 16-hour predictions within RMSE of 2.5 m/s and 48-hour predictions within RMSE of 3.5 m/s.  When applied to the same location of training data, the models generate reliable forecasts for periods up to 22 hours, with the added advantage that the models perform consistently throughout the year ahead horizon, independent of the lead time from the training data. The output of the analysis is highly promising to the wind energy industry in wind forecasting for  locations where historical wind speed data are not available for model building and training. 
\end{abstract}


\end{frontmatter} 

\section{INTRODUCTION} 
As the energy demand multiplies the world over and the conventional energy resources deplete alarmingly, wind energy utilisation has attained greater 
importance from the perspective of sustainable development, as it is renewable, clean and comparatively cost effective. As a result, wind energy has already emerged as an 
essential constituent in the global energy mix and is all set to grow to its maximum potential across countries and regions in the years to come. As wind power is directly proportional to 
the cube of wind speed, even slightest variations in wind speed will significantly affect the power output from wind energy generators. Since wind is a fluctuating resource in respect of availability and speed, precise wind forecasting becomes essential, especially when wind power penetration grows, for the effective management of electricity grid to ensure quality power 
supply. Accurate wind forecasts on different lead time scales help wind farms in real-time grid operations, economic load dispatch planning, reserve requirement decisions, market trading, 
maintenance planning and the like.

Wind forecasting continues to be an area of high research interest owing to its practical relevance in the ever-expanding wind energy industry. Costa et al. \cite{costa2008review} have 
reviewed the research in short-term wind prediction over 30 years, giving attention to forecasting methods, mathematical, statistical and physical models, as well 
as meteorology. Foley et al. \cite{foley2012current} and Okumus et al. \cite{okumus2016current} have conducted an extensive review of the current methods and improvements in the 
field of wind power forecasting. Based on the methodology adopted, wind forecasting models are grouped mainly into physical, statistical, data learning and hybrid models. The physical 
models, which utilize different atmospheric parameters are useful for identifying recurring patterns and making long-term predictions. Statistical models assume that
the wind speed fluctuations are stochastic. However, it has
recently been demonstrated that the underlying dynamics of apparent random-
like fluctuations of wind speed measurements is deterministic, low-dimensional
and chaotic \cite{sreelekshmi2012deterministic,    drisya2014deterministic,drisya2018diverse}. Hybrid models have been developed recently by combining 
different methods such as physical, statistical and machine learning methods to enhance prediction accuracy \cite{meng2016wind,han2017non}. Models based 
on artificial neural network and other data learning techniques have also been gaining increased attention in literature in the recent past. Mohandes et al. \cite{mohandes2004support} have 
compared a support vector regression (SVR) approach for wind speed prediction favourably against a multi-layer perceptron (MLP) for systems with 
orders 1 to 11. Liu et al. \cite{liu2014short} attempted short-term wind speed forecasting using wavelet transform and Support Vector Machines, applying a 
genetic algorithm for parameter optimization. Fugon et al. \cite{fugon2008data} applied linear and non-linear data mining algorithms for the short-term wind power 
forecasting at three distinct wind farm locations in France. Lahouar et al. \cite{lahouar2017hour} tried out Random Forest model for an hour ahead wind power prediction and tested the model with measured wind data and showed good improvement of forecast accuracy compared to classical neural network prediction. Mohandes et al. \cite{mohandes2012spatial} conducted a study within Saudi Arabia to estimate the mean monthly wind speed at certain locations using the historic mean monthly wind speed data from a number of other locations and reported good agreement between estimated and measured monthly mean wind speed values. Browell et al. \cite{browell2018improved} attempted very short-term wind forecasting by incorporating large-scale meteorological information into a vector autoregressive model and showed improved accuracies in forecasting in different case studies conducted in the United Kingdom.

Various machine learning algorithms have been tested successfully to predict wind speed variations. However, almost all reported studies are location specific as training and testing data are sourced from the same location. Apart from that, application of such models have not been tested for time independence as the test data considered are adjacent in time to the training data. In this work, we investigate the possibility of developing time and location independent models based on machine learning algorithms for wind speed forecasting for the wind energy industry. Such models hold practical value for wind energy industry for locations where sufficient past data are not available for model training. The first objective of this study is to investigate the accuracy of models applied to data moving away from the training data set. The second objective is to analyse the accuracy of cross-location predictions in which models trained using data from one location are tested against data from another location.  Proper training of the forecast model with the available data is a crucial factor for its performance. In the case of wind speed time series, it raises the question of deciding the optimum size of past data required to train the model for producing accurate time ahead predictions, as the duration of wind speed time series is significant since the fluctuation characteristics vary remarkably over different seasons. The
theory of mutual information has been applied to estimate the optimum size of preceding data set to be applied in the training of models and testing of forecasts. The study reported here investigates the efficacy of two uniquely trained and tested machine learning models based on (i) Support Vector Machine (SVM) and (ii) Random Forest (RF) algorithms for  wind speed forecasting in the same-location as well as cross-location scenarios, wherein promising results are obtained.

\begin{table}
	\begin{center}
		\begin{tabular}{ |c|c|c|c| } 
			\hline
			Location & Latitude & Longitude \\
			\hline
			L1 & $09^0 45' 30.2''$ & $77^0 10' 41.3''$ \\
			\hline
			L2 & $09^0 59' 09.5''$ & $77^0 11' 50.0''$ \\
			\hline
			L3 &$12^0 01' 32.7''$ & $75^0 20' 32.4''$\\
			\hline
			L4 & $09^0 39' 11.5''$ & $76^0 53' 3.0''$\\
			\hline
			L5 &$10^0 48' 57.7''$ & $76^0 40' 10.3''$\\        
			\hline
		\end{tabular}
		\caption{\label{tab.loc} Geographic coordinates of the wind masts used for wind data collection.
		}
	\end{center}
\end{table}

\begin{figure}
	\centering\includegraphics[width=\columnwidth]{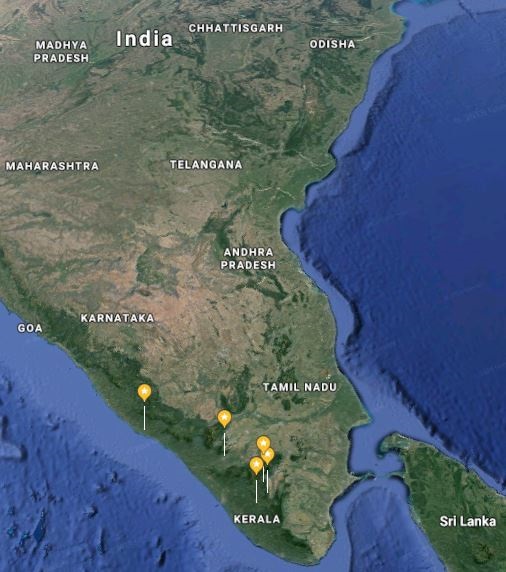}
	\caption{\label{fig.locations} Geographical locations of the wind measuring masts indicated on 3D map (Courtesy: Google Maps).}
\end{figure} 

\section{MACHINE LEARNING MODELS}
Machine learning is a method of improving the performance of a computational software programme by itself with experience. In machine learning, past
experience is fed to the machine as input, and it gives the output as a typical
model that is capable of solving future problems of the same nature. Here,
past experience is collected for the purpose of imparting training. An abstract
target function is determined that well describes the relationship between existing
input and desired output. Subsequently, a machine learning model is selected
to approximate the target function. In the end, a suitable algorithm is used
to build the model from the training examples. In this paper, two different
machine learning models, namely, Support Vector Machine and Random Forest
models have been used for the investigations. The e1071 package in R is used in this research to develop, train and test the models \cite{e1071}.
\subsection{ Support Vector Machine model}
Support Vector Machine (SVM) is a supervised learning method derived from Vapnik’s work on statistical learning theory which was initially used for classification problems and later generalized for regression \cite{cristianini2000introduction,cortes1995support}.  It is an optimization technique of finding a surface which maximizes the margin between two classes based on two main ideas namely, the maximization of distance between the classifying surface and the nearest elements called support vector and the transformation of the input dimension into higher dimensional space using a kernel function. SVM method performs classification tasks by constructing hyperplanes in a multidimensional space that separate cases of different class labels. To construct an optimal hyperplane, SVM employs an iterative training algorithm which is used to minimize an error function. In the case of a linear classification problem, the hyperplane can be expressed as
\begin{eqnarray*}
	y_i(w \times x_i+b) \geq 1-\psi_i
\end{eqnarray*}
where $x_i, y_i \in R^n$   are the training data pairs and $w$ the coefficient vector of classification hyperplane, b the offset of the hyperplane from the origin and $\psi_i$ are the positive slack variables \cite{cortes1995support}. The optimum hyperplane is obtained by solving the minimization problem .
\begin{eqnarray*}
	Minimize \sum_{i=1}^{n} \alpha_i \frac{1}{2}\sum_{i=1}^{n}\sum_{i=1}^{n}n \alpha_i \alpha_j y_iy_j(x_ix_j) \\
	subject \, to \sum_{i=1}^{n}n \alpha_i y_j = 0 and 0 \leq \alpha_i \leq C
\end{eqnarray*}
Where $\alpha_i$ are Lagrange multipliers and C the penalty \cite{samui2008slope}. RBF kernel has been used in our present analysis. The SVM can also be used for regression without sacrificing its main features and it is resistant to overfitting.
\subsection{Random Forest model}
The Random Forest (RF) algorithm is a non-parametric ensemble based learning technique used for both classification and regression \cite{breiman2001random}. The decision tree algorithm works on a set of rules and the possible outcomes to form a tree-like structure. However, such a system is prone to error propagation contributed by an incorrect rule which adds the impurity to the subsequent nodes. Random Forest algorithm eliminates error diffusion process inherent in decision trees by constructing multiple trees. Random samples of given data set are generated and fed to several tree-based learners to form a random forest. Splitting condition for each node in a tree is based only on the randomly selected predictor attributes which lower the error rate by avoiding the correlation among the trees. The successful application of random forest regression algorithm has already been reported in many fields like cheminformatics, speech recognition, bioinformatics, classification and prediction in ecology \cite{svetnik2003random,xu2004random,jiang2004joint}. The random forest regression, which is a non-parametric, captures the functional relationship between dependent and independent variables from the features of the data. From a given data set, algorithm generates a forest of $n$ trees as $\{T_1(X),T_2(X),\hdots, T_n(X)\}$, using an $m$ dimensional vector input $X=(x_1,x_2,\hdots ,x_m)$. Every tree $T_i(X)$  generates an outcome $W_i=T_1(X)$. The average of all individual outputs is considered as the response of random forest. The bagging process of selection with replacement is done both on the samples and the attribute. Normally two third of the data will be the size of bootstrap samples, and the rest is known as out-of-bag samples. The combined effect of bootstrap and attribute bagging helps the algorithm to reduce misclassification error.

\begin{figure}
	\centering\includegraphics[width=\columnwidth]{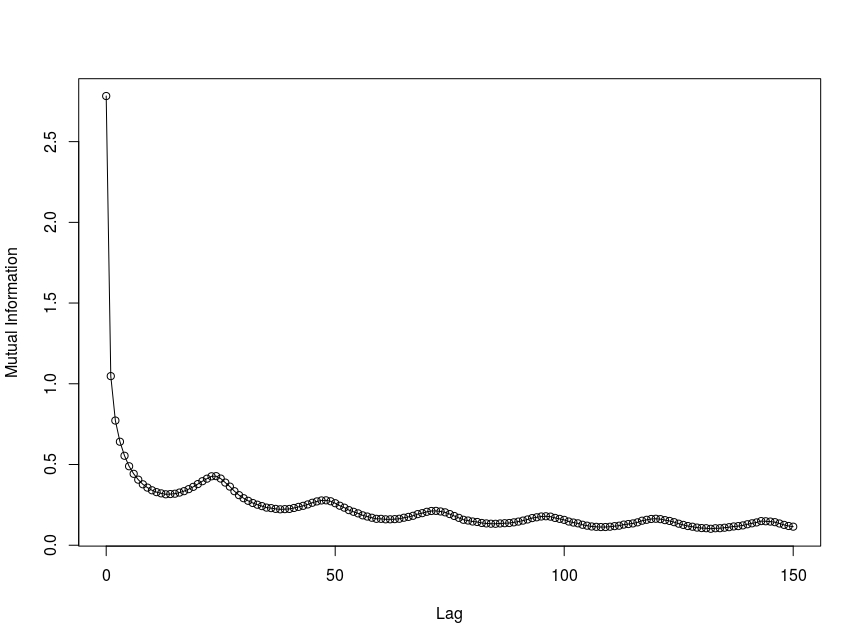}
	\caption{\label{fig.mutual} The mutual information of hourly wind speed time series as function of delay.}
\end{figure}

\begin{figure}
	\centering\includegraphics[width=\columnwidth]{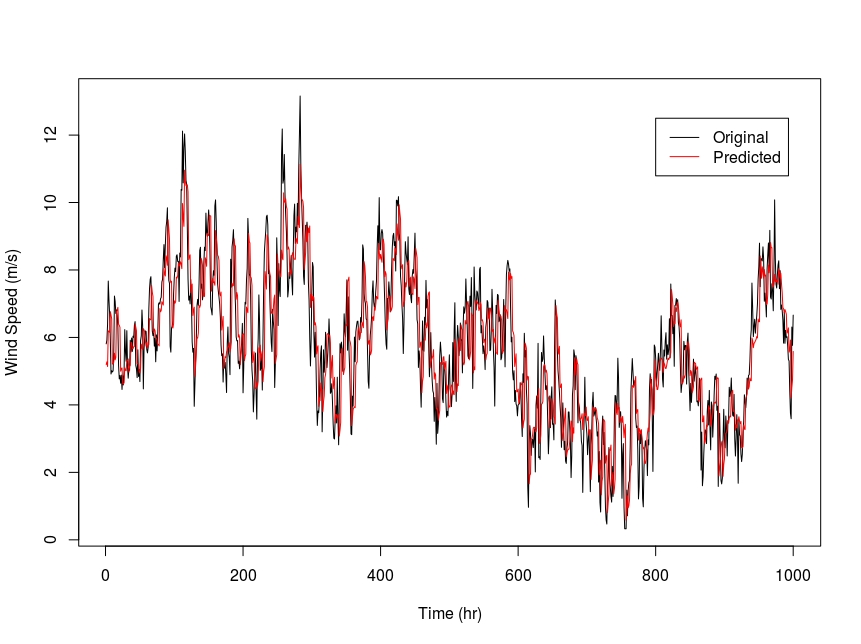}
	\caption{\label{fig.sameloc_3hr_kul} Comparison between the actual and forecast wind speeds with repeated 3 hour ahead predictions for the same-location forecasting for location L1, using SVM model.}
\end{figure}

\section{RESULTS AND DISCUSSIONS}   

The hourly wind speeds at the height of 80 m above the ground level at five
different windy locations situated in the Indian state of Kerala, represented by L1 to L5 as shown in  3D map given in Fig.~\ref{fig.locations} and summarised in Table~\ref{tab.loc}, measured using
wind masts over a continuous period of two years 2012 and 2013 have been utilised for the analysis in this work. These locations are geographically
distributed in such a way that the locations L2, L3, L4 and L5 are at radial
distances 25 km, 321 km, 34 km and 130 km respectively from the location
L1 and the triangular area formed by the locations L1, L2 and L4 comes to 385
square kilometres.  Two different machine learning forecast models built on SVM and RF algorithms have been employed for the analysis. The  wind speed data for
the first year (2012) have been used for training and validating the models, in
order to ensure effective learning of the dynamics of wind flow fluctuations over
a complete seasonal cycle. The wind speed data for the second year (2013) have been used for the testing of forecast results.

\begin{figure}
	\centering\includegraphics[width=\columnwidth]{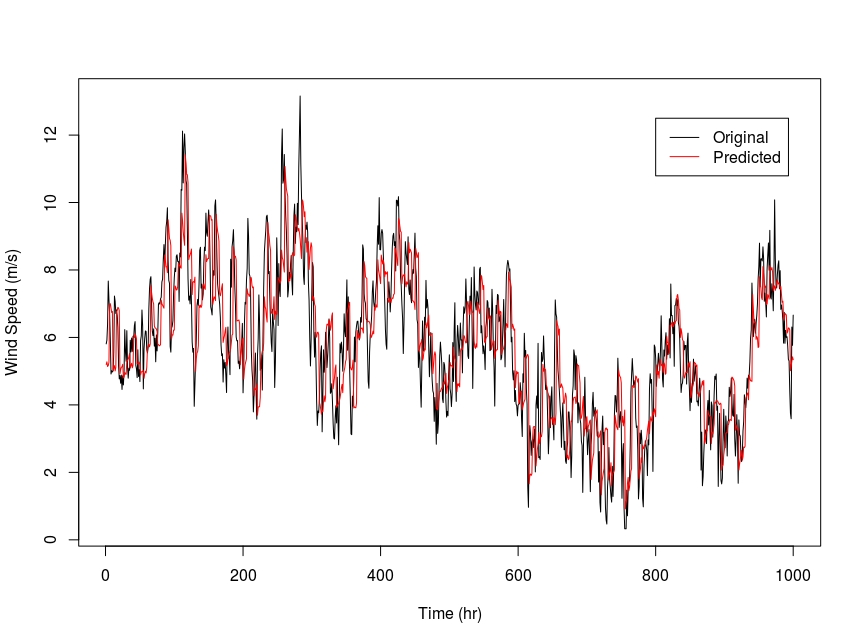}
	\caption{\label{fig.sameloc_5hr_kul} Comparison between the actual and forecast wind speeds with repeated 5 hour ahead predictions for the same-location forecasting for location L1, using SVM model.}
\end{figure}
\begin{figure}
	\centering\includegraphics[width=\columnwidth]{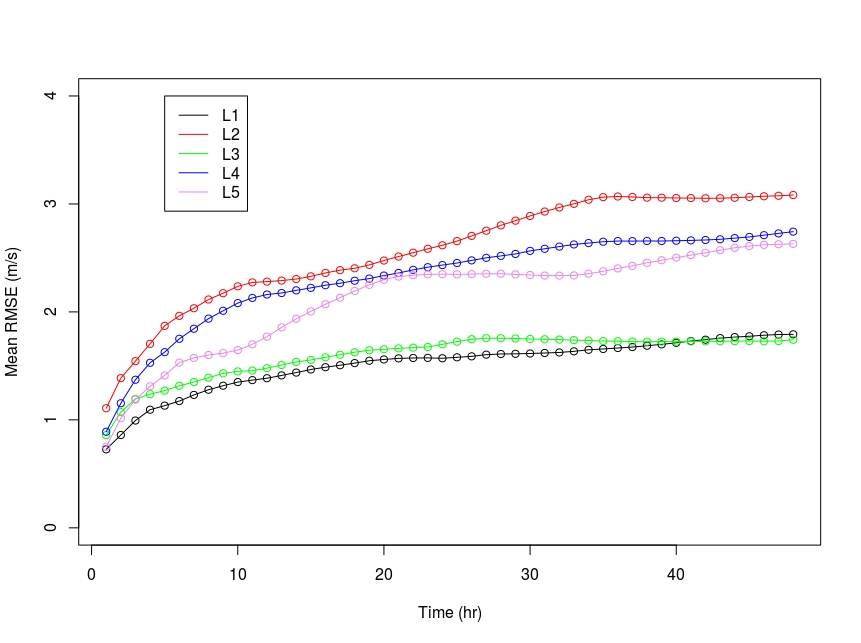}
	\caption{\label{fig.same_svmrmse1} Mean of RMSEs of predictions versus prediction time, for predictions at the locations L1 to L5 using SVM model trained with data from the respective locations. }
\end{figure}

The extent of dependence of a value in a time series on the previous values
can be estimated by the calculation of mutual information between delayed time
series \cite{fraser1986using}. Relying on this, the optimum length of dependency of wind speed
values on the past data has been determined by computing values of mutual
information and the same knowledge made use of in the training of models and
testing of forecasts. The mutual information in the present analysis appears
to become negligible by 72 data points, after which the plot starts
to level-off, as observed in Fig.~\ref{fig.mutual}. It is therefore taken that every wind speed
data point is a function of its previous 72 data points. Hence, the training of SVM 
model using the past data has been done for all the wind speed values in the
training data segment, by inputting 72 continuous measures for each wind speed
value ahead. The same approach has been adopted while inputting the previous set
of data for training the RF models as well. 

In the first stage, same-location predictions (where data from the same location is used in parts for training and testing) have been examined for all the
five locations, for all hours ahead predictions from 1 hour  up
to 48 hours ahead using SVM model. Each of the one step ahead
predictions is generated by inputting the immediately preceding 72 data points
into the trained model. In 2 and higher hour ahead predictions, one step 
ahead predictions are repeated, each time by using the lastly predicted value as
the last of the preceding 72 input points. The predicted time series segments have been compared with actual values along the one year test period, and the deviations analysed using the statistical measure of Root Mean Square Error (RMSE). In the next step,
all possible combinations of cross-location predictions have been experimented,
wherein, SVM model trained with data from one location has been employed to
generate wind speed forecasts at the other four locations by inputting test data
from those four locations respectively. In the subsequent stage, all the above
investigations have been repeated by developing and using the machine learning
model of RF and the results obtained with both the models have
been analysed and compared.


In the typical prediction scenario we input past 72 wind speed data points and obtain one step ahead prediction. Further values are predicted recursively using one step ahead prediction. 
Fig.~\ref{fig.sameloc_3hr_kul} shows a typical 3-hour ahead prediction up to 1000 hours ahead of the training data set at location L1 using SVM model.  Similar plot of 5-hour ahead prediction is given in Fig.~\ref{fig.sameloc_5hr_kul}. In both these plots, the predictions are seen remarkably close to the original measured data, except for some over predictions at peaks.
With the SVM model trained using the data in 2012, we obtained predictions up to 48 hours ahead corresponding to each data point in 2013 by inputting 72 previous values. These predictions were compared with actual data to find the RMSE. Fig.~\ref{fig.same_svmrmse1} shows the mean RMSE versus prediction time averaged over predictions with respect to each data point in 2013 for all the five locations. It may be noted that the RMSE is less than 2 m/s up to 48-hour ahead prediction for locations L1 and L3 whereas the same is almost less than 3 m/s for other locations. However, up to 22-hour ahead, predictions show RMSEs less than 2.5 m/s in all the locations.

\begin{figure}
	\centering\includegraphics[width=\columnwidth]{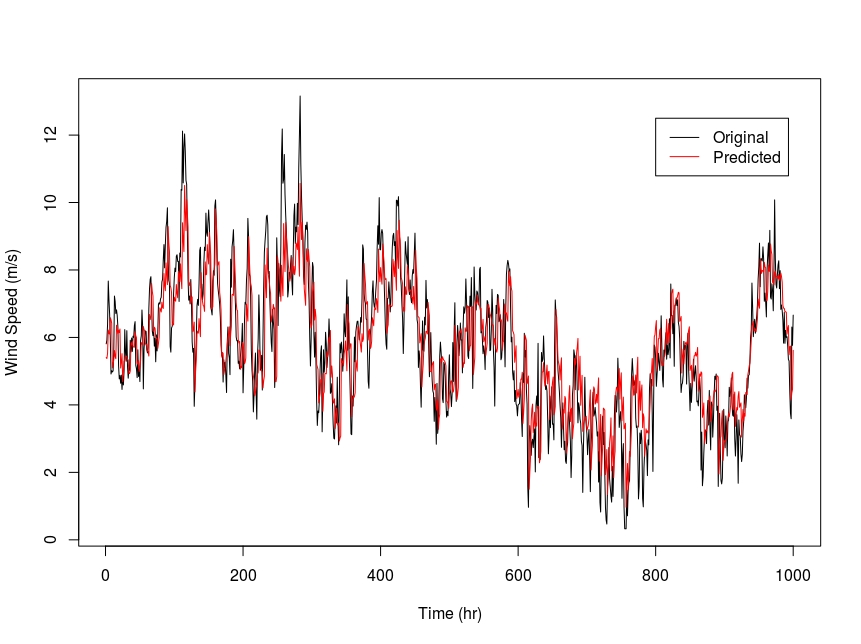}
	\caption{\label{fig.mal_kul_3hr} Comparison between the actual and forecast wind speeds with 3-hour ahead predictions for the cross-location forecasting for location L1, using SVM model trained with data from location L5.}
\end{figure}

\begin{figure}
	\centering\includegraphics[width=\columnwidth]{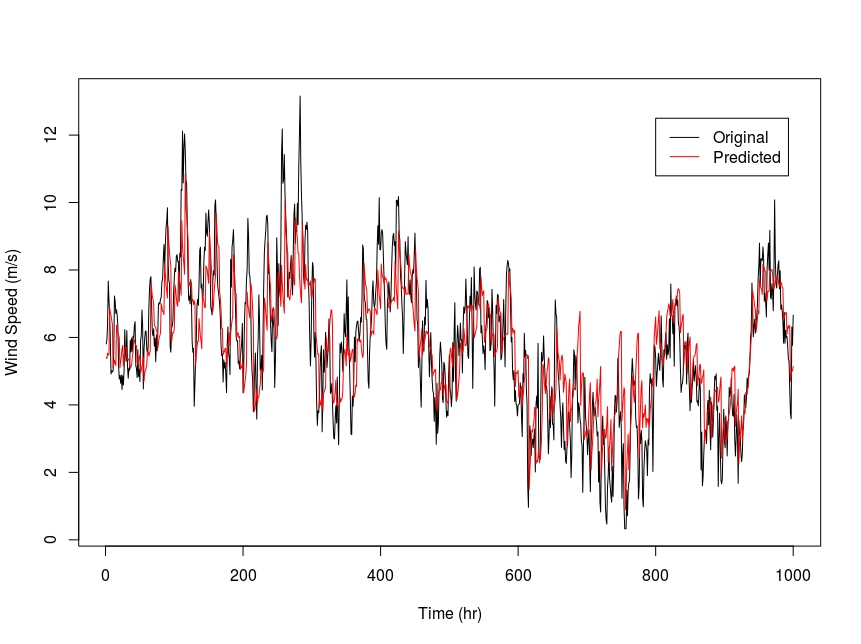}
	\caption{\label{fig.mal_kul_5hr} Comparison between the actual and forecast wind speeds with 5-hour ahead predictions for the cross-location forecasting for location L1, using SVM model trained with data from location L5.}
\end{figure}

\begin{figure}
	\centering\includegraphics[width=\columnwidth]{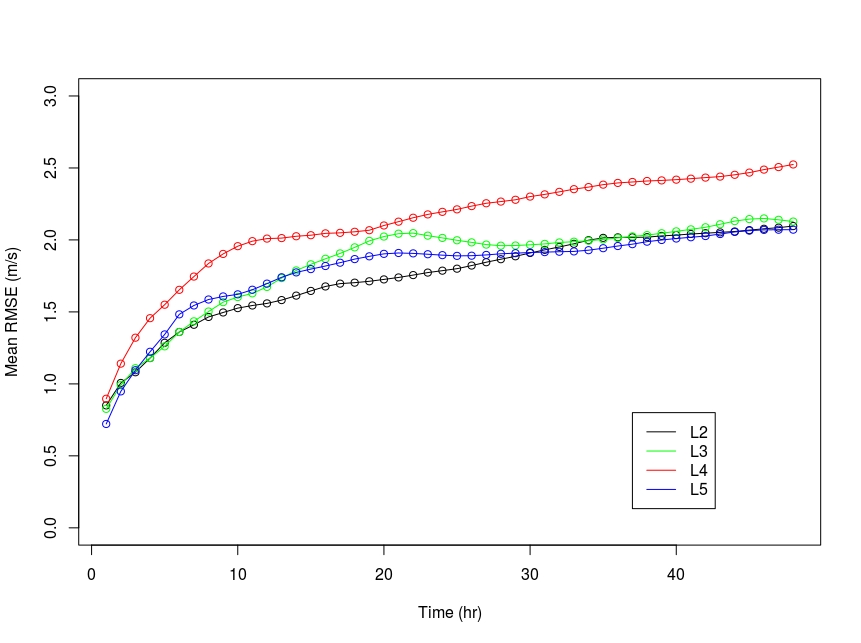}
	\caption{\label{fig.kul_rmse1} Mean of RMSEs of predictions versus prediction time, for cross-location predictions at the locations L2, L3, L4 and L5 using SVM model trained with data from the location L1. }
\end{figure}

\begin{figure}
	\centering\includegraphics[width=\columnwidth]{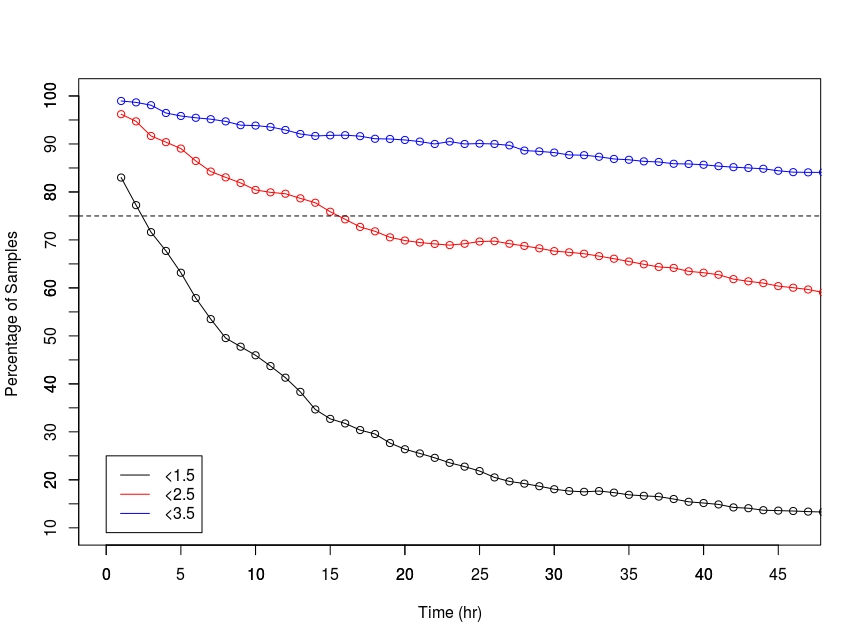}
	\caption{\label{fig.precentage_SVM1} Percentage of cross-location predictions with RMSEs below 1.5 m/s, 2.5 m/s and 3.5 m/s versus prediction time, when using SVM model.}
\end{figure}

\begin{figure}
	\centering\includegraphics[width=\columnwidth]{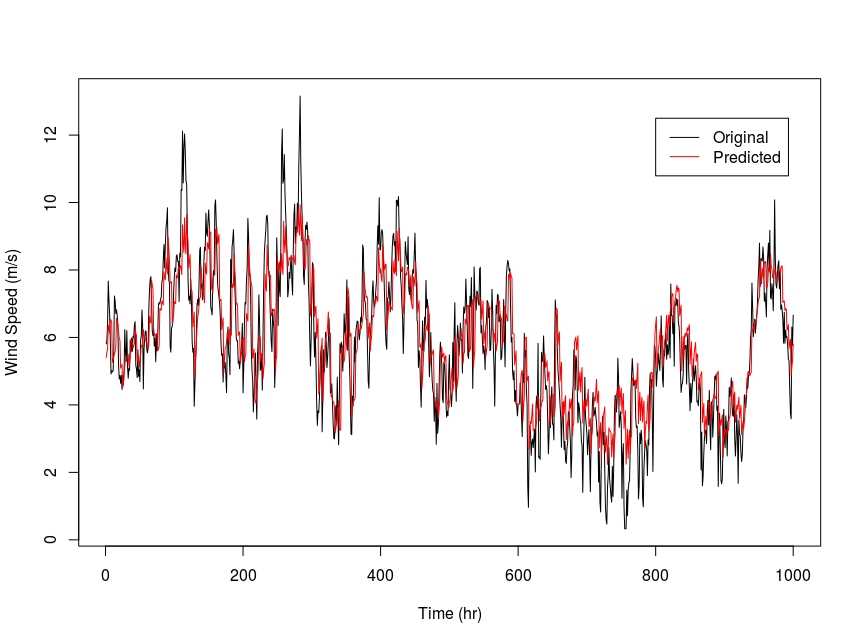}
	\caption{\label{fig.RF_mal_kul_3hr} Comparison between the actual and forecast wind speeds with 3 hour ahead predictions for the cross-location forecasting for location L1, using RF model trained with data from location L5.}
\end{figure}

\begin{figure}
	\centering\includegraphics[width=\columnwidth]{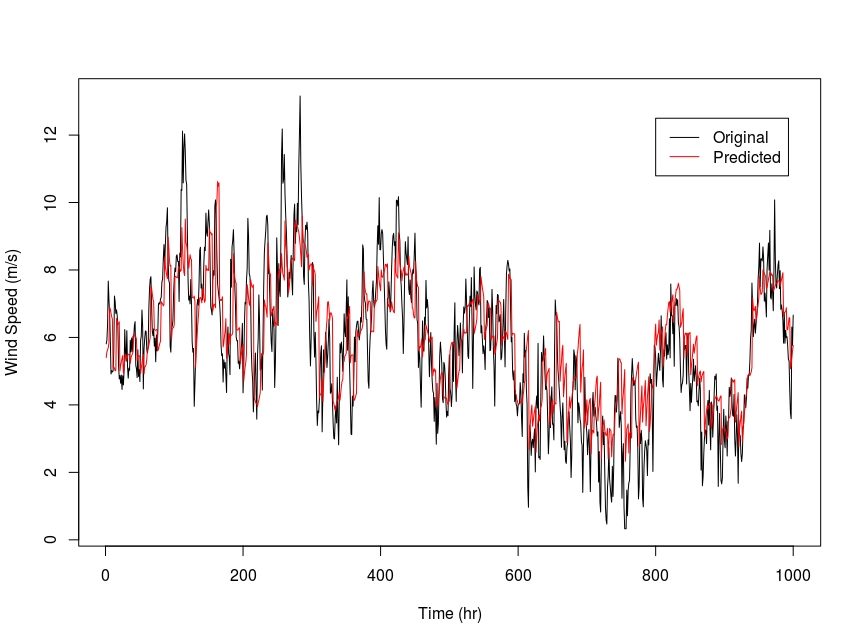}
	\caption{\label{fig.RF_mal_kul_5hr} Comparison between the actual and forecast wind speeds with 5 hour ahead predictions for the cross-location forecasting for location L1, using RF model trained with data from location L5.}
\end{figure}

\begin{figure}
	\centering\includegraphics[width=\columnwidth]{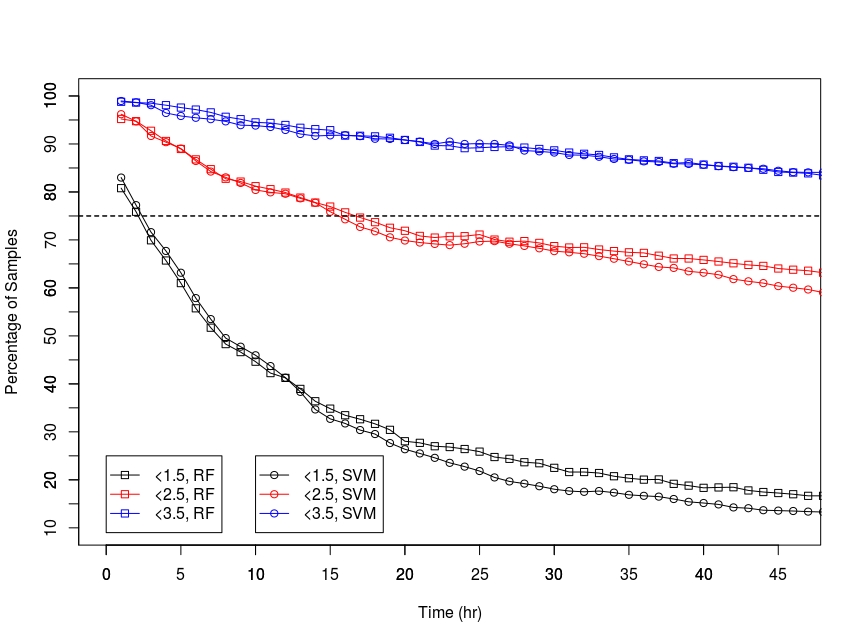}
	\caption{\label{fig.percentage1} Percentage of cross-location predictions with RMSEs below 1.5 m/s, 2.5 m/s and 3.5 m/s versus prediction time, when using SVM model (circular points) and RF model (squared points).}
\end{figure}

As a next step, we trained the model with data of one location of the year 2012 and applied to predict, corresponding to each data point in 2013, up to 48 hours ahead for each other location. Two sample predictions, 3-hour and 5-hour ahead, up to 1000 hours ahead of the time of training data are given in Figs.~\ref{fig.mal_kul_3hr}, \ref{fig.mal_kul_5hr}.  When the above investigation is carried out under the cross-location scenario for the
five locations, a total of 5 prediction scenarios, each with predictions for four
locations using the SVM model trained with the data from the other location is 
available for analyses.  Fig.~\ref{fig.kul_rmse1} depicts one such scenario, wherein predictions are generated for the locations L2 to L5 using the SVM model trained
with the historic data from the location L1. In this case, it can be seen that the mean RMSE is less than 2.5 m/s for up to 48-hour predictions even in the worst case of location L4.  In a similar fashion, we repeated the cross-prediction sequence by selecting one location for modelling and other locations for prediction. In order to have a better understanding of the efficacy of the cross-location predictions, a percentage-wise representation of all predictions throughout the year 2013 for all possible cross-location prediction situations, that show individual RMSE values less than 1.5 m/s, 2.5 m/s and 3.5 m/s are plotted against the prediction time in Fig.~\ref{fig.precentage_SVM1}. 
This plot is useful in two different ways as it helps to determine (i) the length of time ahead predictions that is achievable for a given accuracy of prediction and (ii)  the accuracy levels for a given length of time ahead prediction. For prediction accuracy with RMSE less than 1.5 m/s 75 \% of predictions goes up to 2-hour ahead in time. When the level of prediction accuracy with RMSE is leas than 2.5 m/s, up to 16-hour ahead predictions constitute up to 75\% of the prediction samples. More than 85\% of the 48-hour ahead cross-location predictions are with RMSE less than 3.5 m/s.  In another way of interpreting the above plot is to say that, for example, if 10
hour ahead prediction is considered, around 94\% of prediction samples show
individual RSME values below 3.5 m/s, 80\% of samples show the same value
below 2.5 m/s and 45\% of samples show the same below 1.5 m/s.

In the final phase of the research, the above investigations have been repeated by employing RF model in place of SVM model. The results show
predicted values to match closely with the actual wind time series dynamics
almost as seen as in the case of SVM model, both for same-location as well as cross-location forecasting. The typical cross-location predictions presented in Fig.~\ref{fig.mal_kul_3hr} and Fig.~\ref{fig.mal_kul_5hr} are reproduced in Fig.~\ref{fig.RF_mal_kul_3hr} and Fig.~\ref{fig.RF_mal_kul_5hr} respectively with RF model in place of SVM model for the forecasting, which again show comparable results in terms of wind flow dynamics and forecast accuracy.

In Fig.~\ref{fig.percentage1}, the scenario in Fig.~\ref{fig.precentage_SVM1} for SVM model is reproduced with the corresponding results generated by the RF model overlaid. As can be seen from  Fig.~\ref{fig.percentage1} both models show similar behaviour, especially for lower time ahead predictions. If more stringent levels of forecast accuracy is desired, the SVM model shows
marginal supremacy over the RF model in the shorter time ahead predictions and
vice versa in the longer time ahead predictions.

\section{CONCLUSION}

In this work, we investigated the prospect of employing machine learning based predictive models for the cross-location prediction of wind speed variations. We analysed wind speed data for a period of two years from 5 locations separated by a minimum of 25 km and a maximum of 321 km by using both Support Vector Machine (SVM) and Random Forest (RF) models. The dependency of wind speed on past data has been assessed using the theory of mutual information and the same estimate has been made use of in the intelligent training of models with proper time delay embedding matrix and also in inputting proper length of past data in the testing of forecasts. The results indicate that the models developed and trained here can be effectively used in wind speed forecasting on the same time series at far away points in relation to the training data with respect to time. This time-independent characteristic is helpful in avoiding the need in the usual methods for training of models using the immediate past data every time predictions are attempted. The research further proves that both these models, together with the methods of training and testing followed here, can generate reliable and quality cross-location short-term wind speed forecasts to a duration of 16 to 17 hours from a given point in time across a geographical area as wide as in the present case, with not less than 75\% of such time ahead prediction samples from all along the testing period showing individual RMSE values below 2.5 m/s. In the case of one hour ahead predictions along the one year test period of all the 20 cross-location prediction scenarios from the 5 locations, almost 95\% of such predictions show RMSE values below the same value. The results obtained further show that in cross-location forecasting, the RF model slightly outperforms the SVM model in the longer time ahead predictions when higher levels of accuracies are desired. The promising results obtained in the cross-location forecasting of wind speed point to the possible existence of certain collective characteristics hidden within the surface wind flow dynamics, which deserve to be studied further. From the practical perspective, the research outcome is very promising for supporting the growing wind energy industry, by being able to help develop hardware instruments embedded with trained models for time as well as location independent wind speed forecasting tasks. The cross-location forecast capability makes it possible to predict wind speeds at newly identified locations where sufficient past data are not available for model-building by employing a model trained with historical wind speed data from a different location within that geographical area.

\section*{ACKNOWLEDGEMENT}
The authors acknowledge with gratitude the use of wind data, measured by NIWE (National
Institute of Wind Energy under the Ministry of New and Renewable Energy, Government
of India) for and on behalf of ANERT (Agency for Non-conventional Energy and Rural
Technology under the state government of Kerala in India), in this research work. The authors are also grateful to the campus computing facility of University of Kerala set up under DST-PURDE programme for providing computational facilities. The authors also state that they have no conflicts of interest to declare.   

\section*{REFERENCES}
\bibliographystyle{unsrt}  
\bibliography{mybibfile}

\begin{thebibliography}{10}

\bibitem{costa2008review}
Alexandre Costa, Antonio Crespo, Jorge Navarro, Gil Lizcano, Henrik Madsen, and
  Everaldo Feitosa.
\newblock A review on the young history of the wind power short-term
  prediction.
\newblock {\em Renewable and Sustainable Energy Reviews}, 12(6):1725--1744,
  2008.

\bibitem{foley2012current}
Aoife~M Foley, Paul~G Leahy, Antonino Marvuglia, and Eamon~J McKeogh.
\newblock Current methods and advances in forecasting of wind power generation.
\newblock {\em Renewable Energy}, 37(1):1--8, 2012.

\bibitem{okumus2016current}
Inci Okumus and Ali Dinler.
\newblock Current status of wind energy forecasting and a hybrid method for
  hourly predictions.
\newblock {\em Energy Conversion and Management}, 123:362--371, 2016.

\bibitem{sreelekshmi2012deterministic}
RC~Sreelekshmi, K~Asokan, and K~Satheesh Kumar.
\newblock Deterministic nature of the underlying dynamics of surface wind
  fluctuations.
\newblock {\em Annales Geophysicae-Atmospheres Hydrospheresand Space Sciences},
  30(10):1503, 2012.

\bibitem{drisya2014deterministic}
GV~Drisya, DC~Kiplangat, K~Asokan, and K~Satheesh Kumar.
\newblock Deterministic prediction of surface wind speed variations.
\newblock {\em Ann. Geophys}, 32:1415--1425, 2014.

\bibitem{drisya2018diverse}
GV~Drisya, K~Asokan, and K~Satheesh Kumar.
\newblock Diverse dynamical characteristics across the frequency spectrum of
  wind speed fluctuations.
\newblock {\em Renewable Energy}, 119:540--550, 2018.

\bibitem{meng2016wind}
Anbo Meng, Jiafei Ge, Hao Yin, and Sizhe Chen.
\newblock Wind speed forecasting based on wavelet packet decomposition and
  artificial neural networks trained by crisscross optimization algorithm.
\newblock {\em Energy Conversion and Management}, 114:75--88, 2016.

\bibitem{han2017non}
Qinkai Han, Fanman Meng, Tao Hu, and Fulei Chu.
\newblock Non-parametric hybrid models for wind speed forecasting.
\newblock {\em Energy Conversion and Management}, 148:554--568, 2017.

\bibitem{mohandes2004support}
Mohammad~A Mohandes, Talal~O Halawani, Shafiqur Rehman, and Ahmed~A Hussain.
\newblock Support vector machines for wind speed prediction.
\newblock {\em Renewable Energy}, 29(6):939--947, 2004.

\bibitem{liu2014short}
Da~Liu, Dongxiao Niu, Hui Wang, and Leilei Fan.
\newblock Short-term wind speed forecasting using wavelet transform and support
  vector machines optimized by genetic algorithm.
\newblock {\em Renewable Energy}, 62:592--597, 2014.

\bibitem{fugon2008data}
Lionel Fugon, J{\'e}r{\'e}mie Juban, and Georges Kariniotakis.
\newblock Data mining for wind power forecasting.
\newblock In {\em European Wind Energy Conference \& Exhibition EWEC 2008},
  pages 6--pages. EWEC, 2008.

\bibitem{lahouar2017hour}
A~Lahouar and J~Ben~Hadj Slama.
\newblock Hour-ahead wind power forecast based on random forests.
\newblock {\em Renewable energy}, 109:529--541, 2017.

\bibitem{mohandes2012spatial}
Mohamed~A Mohandes, Shafiqur Rehman, and Syed~Masiur Rahman.
\newblock Spatial estimation of wind speed.
\newblock {\em International Journal of Energy Research}, 36(4):545--552, 2012.

\bibitem{browell2018improved}
Browell J., Drew~D. R., and Philippopoulos K.
\newblock Improved very short term spatio temporal wind forecasting using
  atmospheric regimes.
\newblock {\em Wind Energy}, 0(0):1--12, 2017.

\bibitem{e1071}
David Meyer, Evgenia Dimitriadou, Kurt Hornik, Andreas Weingessel, and
  Friedrich Leisch.
\newblock {\em e1071: Misc Functions of the Department of Statistics,
  Probability Theory Group (Formerly: E1071), TU Wien}, 2017.
\newblock R package version 1.6-8.

\bibitem{cristianini2000introduction}
Nello Cristianini and John Shawe-Taylor.
\newblock {\em An introduction to support vector machines and other
  kernel-based learning methods}.
\newblock Cambridge university press, 2000.

\bibitem{cortes1995support}
Corinna Cortes and Vladimir Vapnik.
\newblock Support-vector networks.
\newblock {\em Machine learning}, 20(3):273--297, 1995.

\bibitem{samui2008slope}
Pijush Samui.
\newblock Slope stability analysis: a support vector machine approach.
\newblock {\em Environmental Geology}, 56(2):255, 2008.

\bibitem{breiman2001random}
Leo Breiman.
\newblock Random forests.
\newblock {\em Machine learning}, 45(1):5--32, 2001.

\bibitem{svetnik2003random}
Vladimir Svetnik, Andy Liaw, Christopher Tong, J~Christopher Culberson,
  Robert~P Sheridan, and Bradley~P Feuston.
\newblock Random forest: a classification and regression tool for compound
  classification and qsar modeling.
\newblock {\em Journal of chemical information and computer sciences},
  43(6):1947--1958, 2003.

\bibitem{xu2004random}
Peng Xu and Frederick Jelinek.
\newblock Random forests in language modeling.
\newblock In {\em Proceedings of the 2004 Conference on Empirical Methods in
  Natural Language Processing}, 2004.

\bibitem{jiang2004joint}
Hongying Jiang, Youping Deng, Huann-Sheng Chen, Lin Tao, Qiuying Sha, Jun Chen,
  Chung-Jui Tsai, and Shuanglin Zhang.
\newblock Joint analysis of two microarray gene-expression data sets to select
  lung adenocarcinoma marker genes.
\newblock {\em BMC bioinformatics}, 5(1):81, 2004.

\bibitem{fraser1986using}
AM~Fraser and HL~Swinney.
\newblock Using mutual information to find independent coordinates for strange
  attractors.
\newblock {\em Phys. Rev. A}, 33:1134--1140, 1986.

\end{thebibliography}

\end{document}